\documentclass{article}
\usepackage[english]{babel}
\usepackage[hyphens]{url}
\usepackage{authblk}
\usepackage[a4paper, margin=1in]{geometry}
\usepackage{amsmath}
\usepackage{graphicx}
\usepackage{subcaption}
\usepackage[colorlinks=true, allcolors=blue]{hyperref}

\title{The Malware as a Service ecosystem}
\author[1,2]{Constantinos Patsakis}
\author[3]{David Arroyo}
\author[2,4]{Fran Casino}
\affil[1]{Department of Informatics, University of Piraeus, 80 Karaoli \& Dimitriou str., 18534 Piraeus, Greece}

\affil[2]{Information Management Systems Institute of Athena Research Centre, Greece}
\affil[3]{Spanish National Research Council (CSIC), Spain}
\affil[4]{Department of Computer Engineering and Mathematics, Universitat Rovira i Virgili, Tarragona, Spain}

\date{}
\begin{document}
\maketitle

\begin{abstract}
The goal of this chapter is to illuminate the operational frameworks, key actors, and significant cybersecurity implications of the Malware as a Service (MaaS) ecosystem. Highlighting the transformation of malware proliferation into a service-oriented model, the chapter discusses how MaaS democratises access to sophisticated cyberattack capabilities, enabling even those with minimal technical knowledge to execute catastrophic cyberattacks. The discussion extends to the roles within the MaaS ecosystem, including malware developers, affiliates, initial access brokers, and the essential infrastructure providers that support these nefarious activities. The study emphasises the profound challenges MaaS poses to traditional cybersecurity defences, rendered ineffective against the constantly evolving and highly adaptable threats generated by MaaS platforms. With the increase in malware sophistication, there is a parallel call for a paradigm shift in defensive strategies, advocating for dynamic analysis, behavioural detection, and the integration of AI and machine learning techniques. By exploring the intricacies of the MaaS ecosystem, including the economic motivations driving its growth and the blurred lines between legitimate service models and cyber crime, the chapter presents a comprehensive overview intended to foster a deeper understanding among researchers and cybersecurity professionals. The ultimate goal is to aid in developing more effective strategies for combating the spread of commoditised malware threats and safeguarding against the increasing accessibility and scalability of cyberattacks facilitated by the MaaS model.
\end{abstract}

\maketitle

\section{Introduction}
The time when malware could be considered a slight annoyance or a minor threat for organisations has long passed. Modern malware is a multimillion-dollar industry and the spear of the underground economy, which, among others, monetises access to compromised machines, harvests credentials, performs denial of service attacks and keeps organisations hostage with ransomware and syphons illicit traffic \cite{10.1145/3199674}. This underground economy is huge and transnational. For 2023, Cybersecurity Ventures estimated the cost of cybercrime to \$8 trillion and to steadily reach \$10.5 trillion by 2025 \cite{cybersecurityventuresCybercrimeCost}. Practically, if this were a country, it would have the third-largest economy in terms of GDP, doubling the size of the fourth nation. While the latter statistic shows the cost of cybercrime and not the income of the cyber criminals, it is worthwhile to see the incomes of these groups. Darknet markets have been well-known for siphoning illicit payments. Hydra market was estimated to have facilitated the exchange of \$5.2 billion \cite{justiceJusticeDepartment}. Similarly, the DarkMarket was estimated at \$140 million \cite{europaDarkMarketWorlds}, while in the takedown of Monopoly Market, the authorities seized  \$53.4 million in cash and cryptocurrencies \cite{europaDarkVendors}, which essentially is only a fragment of what the market actually facilitated. Other marketplaces, e.g., Genesis, are known to sell credentials of millions of users \cite{europaTakedownNotorious}. On the other hand, ransomware loot, as declared by victims, is almost \$300 million \cite{ransomwheRansomwhere}. Nevertheless, this is just the tip of the iceberg.


Although cybercrime is not merely malware, it is often the spear used to disrupt systems and services and is the leading facilitator of many of these actions. Therefore, a study of its ecosystem can provide many insights to researchers and justify other observations, even at a technical level. To do so, we must consider the modern status quo and the evolution of malware to malware as a service (MaaS). Indeed, MaaS represents a pivotal shift in the landscape of cyber threats, transforming malware proliferation into a commoditised industry. This evolution mirrors broader trends in the digital economy towards service-based models, where complex tools and services are made available through an accessible, subscription-based framework. In the cybersecurity domain for many years, we have witnessed the democratisation of access to many mature cybersecurity tools for legitimate purposes, e.g., penetration testing. In the same sense, MaaS democratises access to malware capabilities, lowering the barrier to entry for cybercriminals by enabling even those with minimal technical expertise or no infrastructure to launch sophisticated and catastrophic attacks. 

At its core, MaaS is a testament to the ingenuity and adaptability of threat actors within the cyber landscape. By leveraging the service-oriented model, malware authors and distributors have created a highly efficient and scalable method for deploying malware. This ecosystem not only includes the developers who create and maintain the malware, but also encompasses affiliates who distribute the malware and numerous other roles that support this shadow economy, including, but not limited to, payment processors, marketing specialists, and customer support agents. The MaaS model has facilitated the proliferation of a wide array of threats, from ransomware to trojans, botnets, and beyond, each tailored to specific criminal objectives through an array of payment models, including subscription-based services and pay-per-use.

The emergence of MaaS has profound implications for cybersecurity defences. For instance, traditional defence mechanisms that rely on signature-based detection or static analysis are rendered useless against the polymorphic and metamorphic malware variants spawned by MaaS platforms. In parallel, the increasing sophistication of mobile malware is also another threat challenging well-established detection tools and methodologies \cite{cccOperationTriangulation,zimperium2023Global}. These challenges necessitate a parallel paradigm shift in defensive strategies, encompassing robust, resilient, and adaptive systems. Therefore, dynamic analysis, behavioural detection, and incorporating artificial intelligence and machine learning techniques augmented by cognitive security \cite{ANDRADE2019102352} are crucial to identify and neutralise such threats preemptively. As a result, attack vectors associated with the deployment and use of artificial intelligence, namely data poisoning \cite{10.1145/3585385}, trojan attacks \cite{9581257} and other variants, should be considered. 

This chapter provides a comprehensive overview of the MaaS ecosystem, identifying key trends, challenges, and future directions in the fight against commoditised cyber threats. By exploring the actors, mechanisms, and impacts of MaaS, we aim to foster a deeper understanding of this phenomenon and shed light on this illicit ecosystem.

\section{The ecosystem}
\begin{figure}
    \centering
    \includegraphics[width=.7\textwidth]{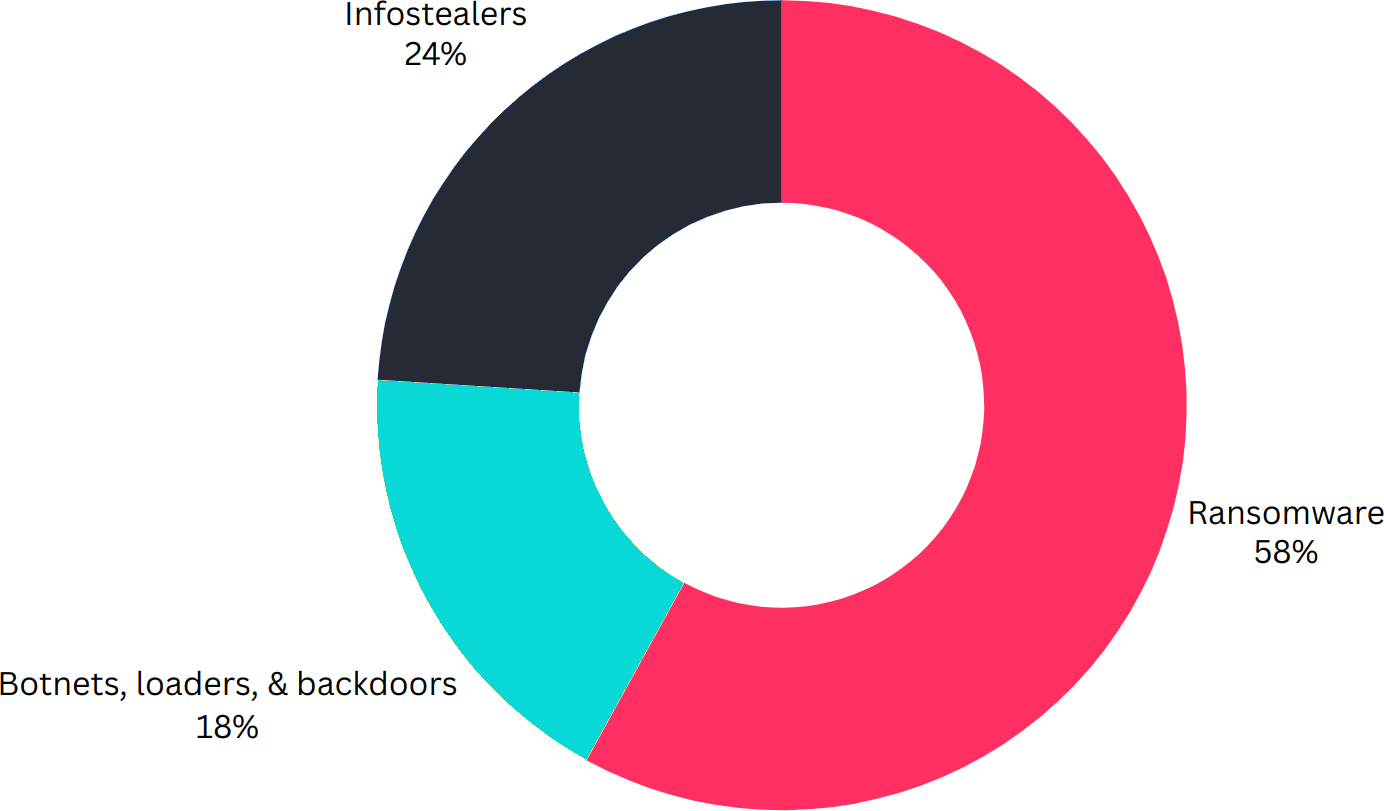}
    \caption{Malware families distributed under the MaaS model between 2015 and 2022. Adapted from \cite{kasper}}
    \label{fig:maas_share}
\end{figure}

The MaaS ecosystem has been driven by the evolution of ransomware, as it is the main powerhouse in terms of malware families distributed under this model. The collaboration and financial exchanges between threat actors are by no means a new development. However, the direct monetisation of ransomware and the delegation of duties in these groups led to the rise of ransomware as a service (RaaS) \cite{meland2020ransomware,karapapas2020ransomware}, and fueled the transition towards MaaS. Indeed, according to Kaspersky \cite{kasper}, the large number of malware families distributed under the MaaS model between 2015 and 2022 are dominated by ransomware. Then, we have infostealers, followed by botnets, loaders, and backdoors, as seen in Figure \ref{fig:maas_share}.

\begin{figure}[th!]
    \centering
    \includegraphics[width=.8\textwidth]{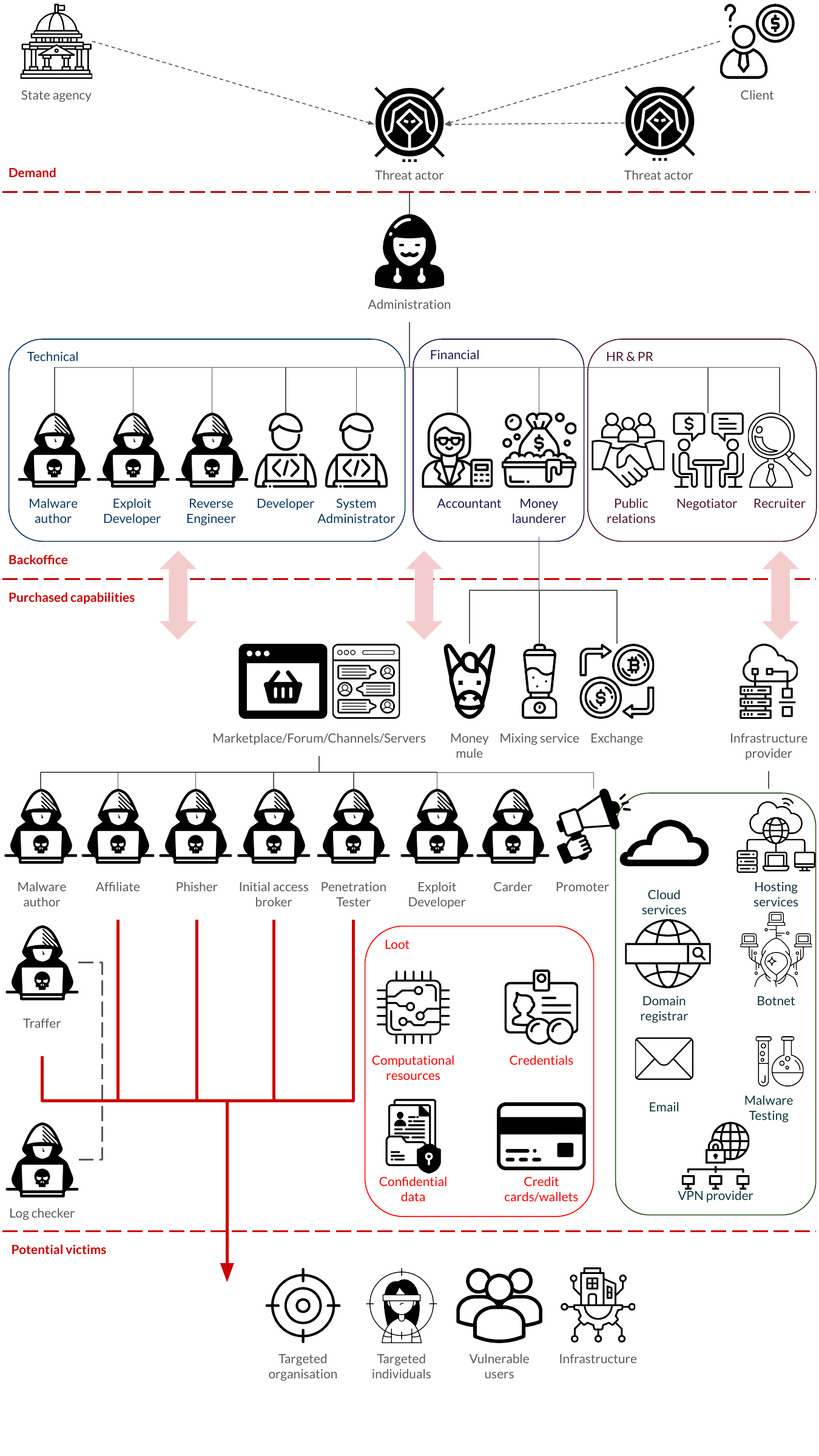}
    \caption{An overview of the MaaS ecosystem}
    \label{fig:maas_ecosystem}
\end{figure}

An overview of the MaaS ecosystem is illustrated in Figure \ref{fig:maas_ecosystem}. Conceptually, we split the ecosystem into threat actors, their backoffices, the purchased capabilities, and potential victims.

On the top layer of the ecosystem, we have the \emph{demand}, practically the entities that request the services of MaaS. Essentially, we have the threat actors themselves or others who may request malware services from their peers. Nevertheless, we have \emph{state agencies} in the case of advanced
persistent threat (APT) groups (see below) who may sponsor attacks on other nations or organisations, but also \emph{clients}, that is, individuals or organisations that want to sabotage or survey their peers. The latter spans from corporate and military espionage to the surveillance of political rivals. 

\subsection{Threat actors}
The threat actors are on the top layer of the ecosystem. They can be categorised in multiple ways, depending on their skills, motivation, target, and modus operandi. 

In general, we have cybercriminals, individuals, and groups that engage in criminal activities for financial, political, or reputation gain. Their methods vary, including, but not limited to, deploying malware, phishing, ransomware attacks, identity theft, and financial fraud. They can operate individually or as part of organised crime groups, regardless of their motives. However, their activities reside in the cyber layer; therefore, they perform cyber crimes or facilitate other crimes in the physical layer, e.g., drugs, trafficking, and money laundering. 

Regarding skills, \textit{Script Kiddies} are inexperienced hackers who use existing computer scripts or tools to penetrate systems, networks, and websites. They typically lack the expertise to write their own exploits and are motivated by a desire for recognition or the thrill of the challenge rather than financial gain or political motives.

\begin{figure}[th]
    \centering
    \begin{subfigure}{0.48\textwidth}
    \includegraphics[width=\textwidth]{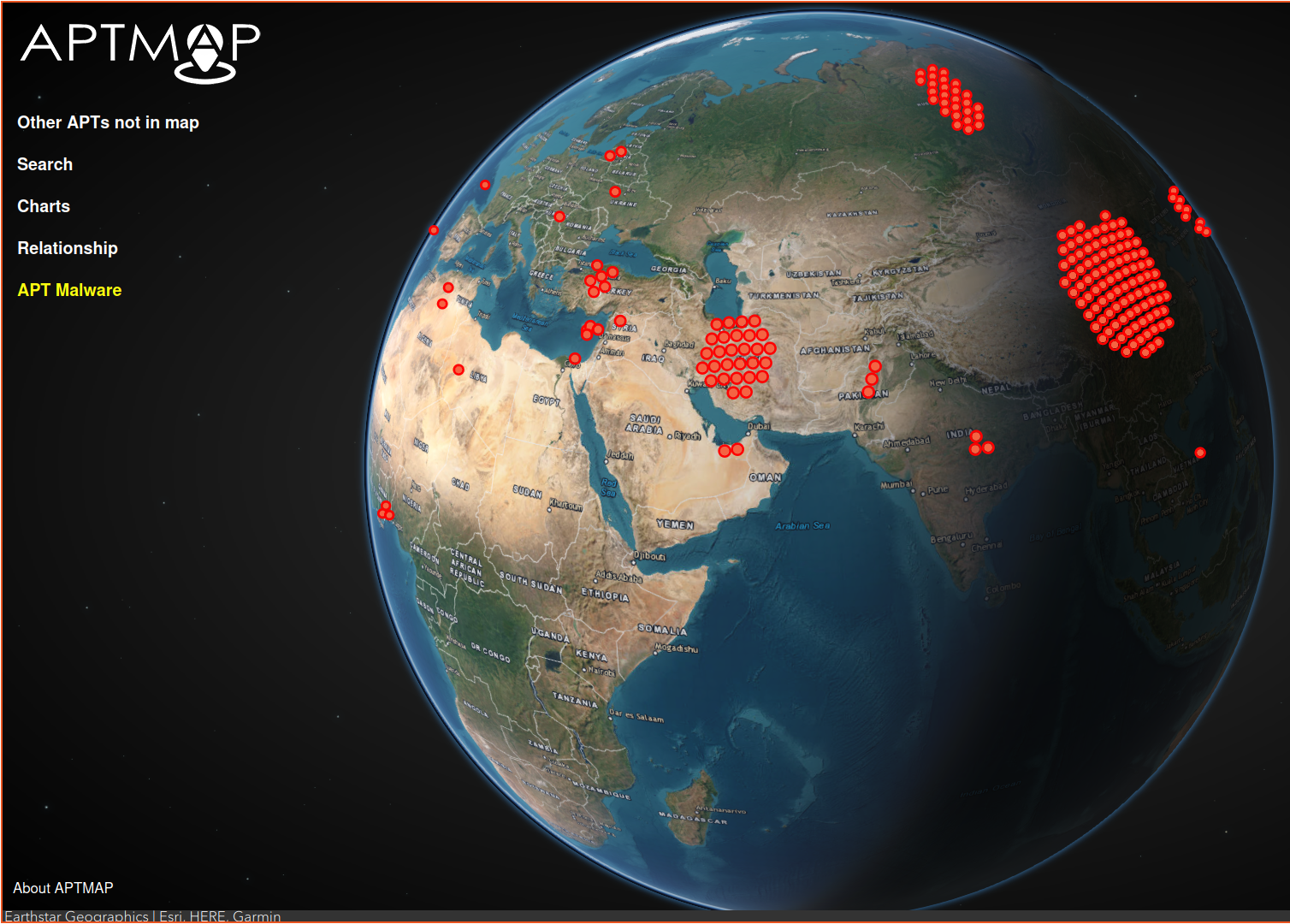}
    \caption{Source: Andrea Cristaldi’s APT Map}
    \end{subfigure}
    \begin{subfigure}{0.48\textwidth}
    \includegraphics[width=\textwidth]{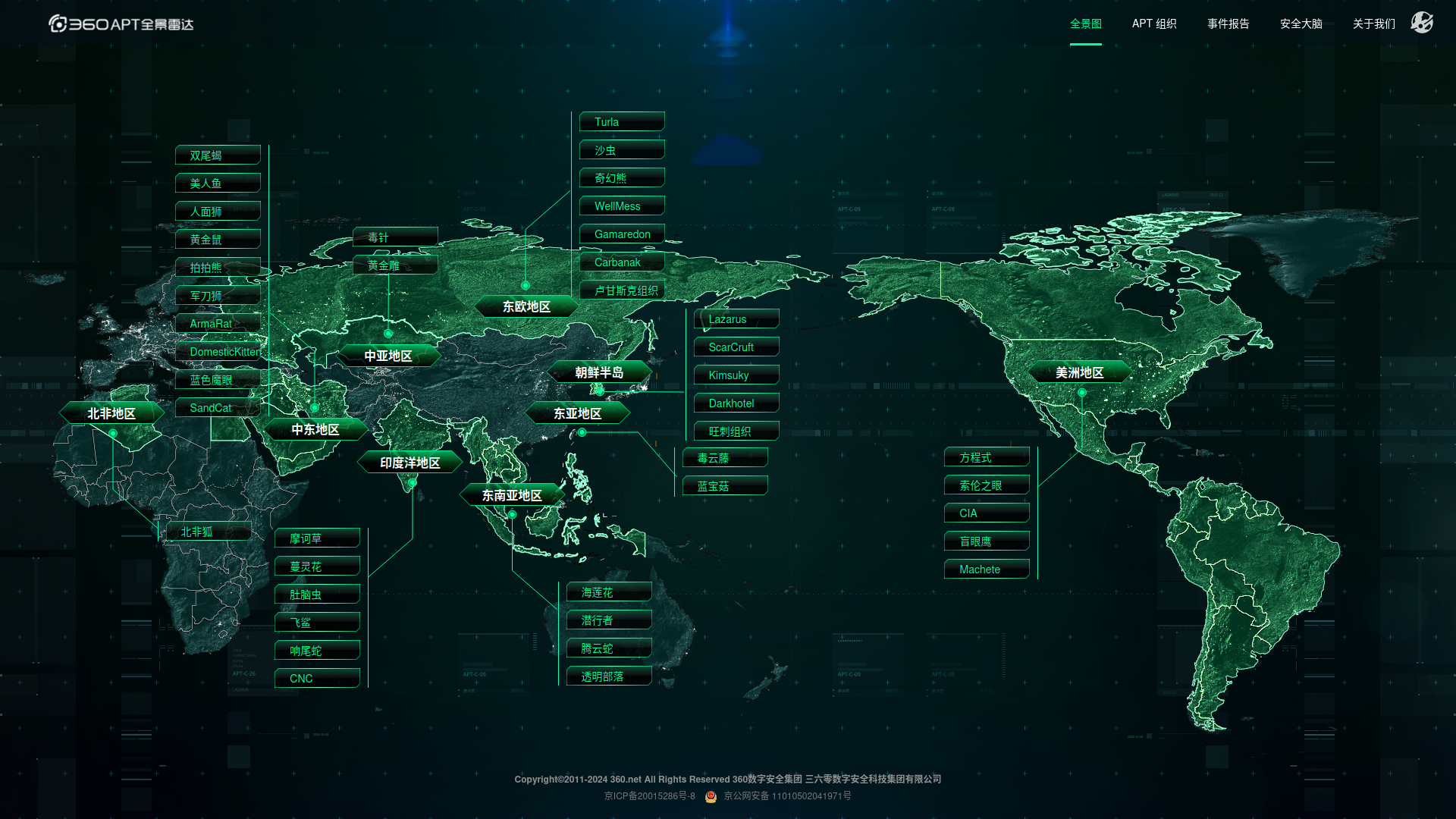}
    \caption{Source: \url{https://apt.360.net/}}
    \end{subfigure}
    \caption{The origin of APT groups.}
    \label{fig:mapapts}
\end{figure}

On the other side of the skill spectrum, we have the APT groups. According to NIST \cite{joint2011sp}, an APT is defined as: "\textit{An adversary that possesses sophisticated levels of expertise and significant resources that allow it to create opportunities to achieve its objectives by using multiple attack vectors (e.g., cyber, physical, and deception)}". As illustrated in the maps in Figure \ref{fig:mapapts}, APTs are primarily based in the East, especially in Russia, China, North Korea, and Iran, according to both Western and Eastern sources. These countries are, in turn, the origin of sophisticated cyberattacks using generative AI \cite{generative_apts}, highlighting potential trends in APTs. Nevertheless, attribution is a thorny topic in APT investigations due to the false-flag operations \cite{skopik2020under}.

Tensions clearly escalated after the Russian invasion of Ukraine, producing an increase in the activity of strategically motivated (S-APTs) such as state-sponsored threat actors. In fact, ENISA stresses that “\textit{geopolitical situations particularly the Russian invasion of Ukraine, have acted as a game changer over the reporting period for the global cyber domain}”\footnote{\url{https://www.enisa.europa.eu/news/volatile-geopolitics-shake-the-trends-of-the-2022-cybersecurity-threat-landscape}}. APT are specialists in espionage and sabotage \cite{Mandiant,talosintelligenceLazarusTale,securelistHelloName} and target-specific infrastructures \cite{langner2011stuxnet}. Therefore, there are numerous reports during the war in Ukraine. For instance, Sandworm compromised eleven telecommunication service providers in Ukraine \cite{certCERTUA} while APT28 launched an attack against a critical power infrastructure facility in Ukraine \cite{certCERTUA2}. 

The geopolitical turmoil, see also Chapter \ref{chap:geo}, magnifies cyber warfare. The latter implies orchestrating long-term disinformation campaigns paired with incoming attacks expected to fearfully exploit unpatched vulnerabilities affecting public and private organisations and critical infrastructures across the world. Since APT groups may not be financially motivated, supply chains have been targeted and are expected to suffer more attacks, as recently outlined by ENISA\footnote{\url{https://www.enisa.europa.eu/publications/threat-landscape-for-supply-chain-attacks}}. In fact, as ENISA reports, more than half of these attacks were attributed to well-known APT groups. Nonetheless, operationally motivated (O-APTs) APTs work in parallel or conjunction with S-APTs, making it difficult to attribute specific operations to achieving broader strategic objectives. Analysing APT groups' Tactics, Techniques, and Procedures (TTPs) and cyber threat intelligence can disrupt their campaigns and give insight into new developments in malware and zero-day attacks. Although the main analysis and characterisation of APT activities are focused on malicious software, in the current context it is necessary to consider how APT can also leverage hardware supply chains, e.g., Slingshot APT \cite{thornton2019politics}.

As stated above, currently state-sponsored actors, individuals or groups directed, funded, or supported by national governments are used to conduct espionage, sabotage, or influence operations against other nations, organisations, or individuals. The fact that these actors are state-sponsored implies that the chances of their prosecution are minimal, yet 
they possess advanced capabilities and infrastructure. Moreover, such actors do not aim at financial gain. Thus, they may bring an organisation to its knees or cause havoc. Moreover, if the goal is espionage, they may try to infiltrate numerous networks and keep a very low profile with as much as possible prolonged duration. Nevertheless, this does not mean that espionage cannot be linked with monetisation, e.g., ransomware attacks \cite{cybereasonPowerLessTrojan}. 

Since one of the primary focus of such groups is espionage and they use advanced methods to exfiltrate information, there has been a lot of research in the past few years on data exfiltration from air-gap systems and the development of specialised malware \cite{guri2018bridgeware}. The fact that these systems are air-gapped and there is no network connection to exfiltrate the data implies using covert channels and exploiting hardware features. Although the latter leads to slow transmission rates, such strategies are difficult to discover. Therefore, one may use powerlines \cite{guri2019powerhammer}, security cameras \cite{guri2019air}, cooling fans \cite{GURI2020101721}, or even abuse DDR SDRAM buses to generate electromagnetic emissions in the 2.4 GHz Wi-Fi bands \cite{guri2022air}, to name a few.

Another actor who is not financially motivated is a hacktivist. Hacktivism is a form of civil disobedience, leveraging cyber tools to protest or take direct action against political decisions, organisations, or nations. Therefore, hacktivism is a cybercrime from individuals or civil groups for a social or political cause. Consequently, it is not state-sponsored and may target the state from which the hacktivists come from. To this end, hacktivists aim to promote their agendas or protest against organisations, governments, or individuals by disrupting services, defacing websites, or leaking confidential information.

On the counterpart of hacktivists, in terms of motives, there are terrorist groups. Indeed, some terrorist organisations have also initiated cyber operations to further their goals. This can include disrupting critical infrastructure, spreading propaganda, recruiting members, siphoning financial transactions or conducting cybercrimes to fund their activities.

While the above describes illegal entities per se, legal entities may also engage in unlawful actions. For instance, businesses may engage in corporate espionage to gain a competitive advantage. This can involve cyber attacks to steal intellectual property or sensitive business data or disrupt a competitor's operations. In this sense, they may engage with individuals within an organisation, such as employees, contractors, or business partners, to misuse their access to harm the target organisation. Their actions can be malicious (e.g., intending to steal data, sabotage systems), or unintentional, resulting from negligence or lack of awareness. 

Of particular interest are three threat actors in this ecosystem: phishers, operators of banking trojans, and ransomware gangs. Phishers are individuals or groups specialising in phishing operations aiming to steal sensitive information such as login credentials and financial information by masquerading as trustworthy entities in electronic communications. Such entities may often be used to serve malware droppers, typically in the form of trojanised MS Office documents \cite{koutsokostas2022invoice} and PDFs, to numerous or targeted individuals. Thus, phishers provide the first foothold to an organisation either by delivering malware or providing credentials for accessing the target's hosts. As a side effect of data breaches in ransomware attacks, lateral phishing \cite{ho2019detecting} is another type of threat of increasing relevance due to the capabilities to create deep fakes by using the datasets associated with email accounts \cite{bethany2024large}. 

Due to the rise of online banking, there is a rise in banking trojans. As the name suggests, it is a specific type of trojan that pretend to be legitimate software but harvest credentials for online banking, cryptocurrency, and other financial services. To do so, they can be equipped with several templates to mimic the targeted institutions \cite{bb,godf} or utilise various methods to trick users into disabling authentication mechanisms \cite{chameleon}. To distribute the malware, the operators use the MaaS ecosystem to increase their outreach and victims. 

Ransomware groups are cybercriminal groups focused on deploying ransomware to extort money from their victims. They encrypt the victim's data and demand a ransom for the decryption key. While encryption could always be abused to encrypt the victim's hosts, it is only in the last decade that this form of crime has flourished and become an industry. The reason is that the ransom can currently be transferred to the criminal with many privacy guarantees and without the intervention of traditional financial institutions. The introduction of Bitcoin and other cryptocurrencies that enable privacy-preserving protocols such as Monero or the use of mixers, has enabled cybercriminals to perform arbitrary financial transactions worldwide without the control of financial institutions. Depending on the modus operandi of the ransomware group, we may have automated and human-operated ransomware \cite{microsoftDefineRansomware}. In the first form, ransomware can spread like a virus or a worm that infects devices it reaches. However, in human-operated ransomware attacks, cybercriminals compromise the hosts of an organisation (physical or cloud) and try to elevate their privileges by examining each host, using various exploits, brute-forcing credentials, etc., but also establish persistence to avoid being locked out of the system. Then, they perform lateral movement trying to penetrate all the reachable hosts, prioritising targets, e.g., servers. To guarantee that their victims will pay, they also destroy all the backup mechanisms that they will find in their victims' networks. At the end of the attack, they typically perform a double extortion to maximise the chances of making their victim pay the ransom. They would exfiltrate confidential data to remote storage so that the ransom they request is to provide the decryption key and not to publish the confidential data. In this way, even if the victim has offline backups and can restore its data and operations, it may pay to avoid leaking confidential data. Note that ransomware groups may also employ additional modes of extortion to maximise their chances of being paid the ransom \cite{trendmicroRansomwareDouble}.

From the above, it is clear that automated ransomware will perform a broad targeting and opportunistic attack, trying to quickly encrypt whatever it finds, following a scripted scenario. Nevertheless, human-operated ransomware tries to maximise the impact and cause visible business disruption by targeting critical assets and infrastructure of the victim. Moreover, the use of humans allows the attack to adapt, as now the operator can trace the valuable assets, but also identify countermeasures while the defender applies them.

\subsection{The backoffice}
As discussed, modern cybercriminals operate in the form of organisations; therefore, in the context of MaaS, we can consider a backoffice that performs different and discrete tasks. We categorise the delegation of duties to \emph{technical}, \emph{financial}, and \emph{human resources/public relations}. We consider that a person or a small group of individuals lead the threat group, and they form the \emph{administration} team. They are in charge of management, making strategic decisions on goals, victims, collaborations, acquisition of infrastructure and resources, and money allocation.  

\subsubsection{Technical personnel}
The primary role in the backoffice is that of the \emph{malware author}. As the name suggests, this is the individual or a group of individuals who write the source code for malware. This individual can be a formal member of a threat actor or someone who sells it on a marketplace or advertises it on a forum. The goal of the malware author is to write the malware, but also to equip it with different features and capabilities during a campaign, and to update and maintain it as the malware may have to adapt to, e.g. evade detection, and fix technical issues. A critical difference between the malware authors belonging to the threat group and the external ones is usually access to the source code and the provided support. However, this will be discussed in the following paragraphs.

Depending on the threat actor, technical personnel can be equipped with \emph{exploit developers}, a capacity which can also be purchased. The goal of exploit developers is to write exploits and bypasses for security mechanisms to facilitate the operations of threat actors. To this end, the exploit developers can follow one or more strategies, such as developing their own exploits, acquiring published exploits to their needs or purchasing them. Next, they will convert such exploits into toolkits or modules of specific malware. In this way, either the malware has integrated the functionality to exploit a particular vulnerability or the exploit is provided in the form of a tool that is offered to the team that performs the attacks so that they can, e.g., elevate their privileges, execute arbitrary code, or perform a denial of service attack.

Another part of the technical stuff of a threat actor can be \emph{reverse engineers}. Their goal is to understand the internals of binaries and firmware, collaborate with malware authors, and exploit developers to provide insight into the capacities of their target victims and their security mechanisms. Often, reverse engineers would purchase and install security solutions on their premises, try to reverse them to understand how they work, and develop methods to bypass them. Similarly, they would look for hard-coded credentials in software and firmware, a long-standing issue for many developers.

Beyond the above, the threat actor is expected to have \emph{developers} to perform various tasks. For instance, they may develop platforms for internal communication, payment tools, monitoring and managing the group and its payments. Similarly, they may create and update webpages, e.g., the announcement pages of ransomware victims or the pages for the negotiations with the victims in the case of ransomware attacks.

Finally, the groups have system administrators to manage the infrastructure used by the threat actor. Modern malware groups use hundreds of servers on their backend with multiple tiers to maintain their infrastructure, be robust against takedowns, store exfiltrated information, launch attacks with the necessary bandwidth, prevent authorities from identifying them, but also manage an army of bots \cite{lumenEmotetIlluminated,teamcymruTrackingBokBot,thehackernewsQakBotMalware}. Evidently, this infrastructure needs skilled system engineers who can quickly adapt among various infrastructures and are security savvy to avoid leakages, but they must also be on call for immediate actions. 

\subsubsection{Financial personnel}

The first role of financial personnel is the \emph{accountant}. The primary goal of the accountant is to keep track of all the payments that have to be made and all the incoming payments and keep a precise balance. While this part is relatively straightforward, this role has been extended over the past few years, primarily due to the MaaS model, which has significantly increased monetary exchanges between different entities. Nevertheless, the MaaS model implies another critical responsibility of this role: assessing the victim's financial profile. More precisely, accountants, in the case of, e.g., ransomware gangs, have to assess the victim organisation's profile to estimate how much should be asked from the victim as ransom to maximise the probability of being paid and the profit. To this end, the accountant must go through the balance sheets of the victim, exfiltrated financial documents, but also banking transactions from scanned documents, signed contracts, and email conversations to assess how much the victim can pay. Note that organisations have to pay fees for cyber insurance based on their income, industry sector, size, the type and amount of sensitive data an organisation processes and stores, existing cybersecurity measures, etc. An accountant can collect this information by assessing the documents or extracting it from payment slips, contracts, etc.

However, incoming and outgoing payments must be made in a way that avoids tractability from the authorities. As a result, threat actors need people who will launder their payments and income. To this end, the use of money mules is a widespread tactic. Meanwhile, the introduction of cryptocurrencies has dramatically facilitated this, as payments can be performed instantly without intermediaries and controlled by third-party entities. As a result, cryptocurrencies are being abused for various illicit payments, especially cybercrime \cite{10012354,chainreport,elliptic}. Actually, the evolution of the so-called decentralized finance makes it even more difficult to conduct cyber attribution, as new techniques to conceal illegal transactions and perform money laundering can be fulfilled by means of cross-chain bridges, flash loans, and other means \cite{zhou2023sok,carpentier2023mapping}. Therefore, the \emph{money launderer} collaborates closely with the accountant to fuse the payments to money mules or launder cryptocurrencies through mixers and tumblers and then withdraw the money from exchanges. As a response, there is a lot of research on tracking transactions on blockchains and grouping them into entities \cite{reid2012analysis,ron2013quantitative,Haslhofer2021a,CACM-Ransomware}, which can be facilitated by operational errors, greed, and vanity \cite{patsakis2023cashing}.

\subsubsection{HR and public relations}
\begin{figure}[th]
    \centering
    \includegraphics[width=\textwidth]{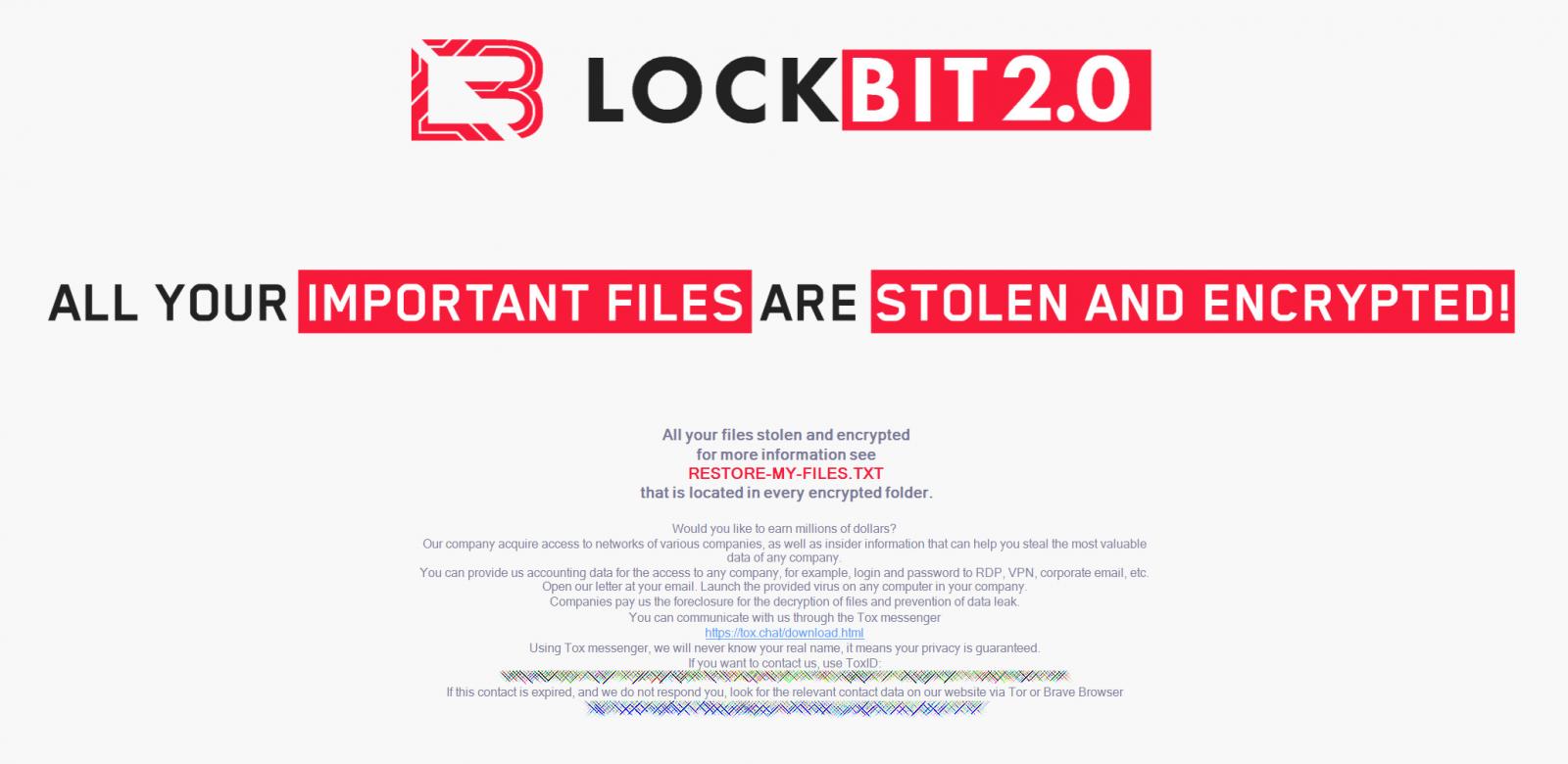}
    \caption{LockBit trying to lure insiders to breach corporate networks.}
    \label{fig:lockbit_insiders}
\end{figure}

The final group of the backoffice deals with the human management of the team and interactions. Therefore, some people are focused on recruiting people. Consequently, they scout people and profiles on forums and social networks, perform background checks, and determine their fitness regarding skills to the group's needs. Nevertheless, over the past few years, there have been many radical approaches to recruitment processes. For example, Fin7, a notorious APT group from Russia, posed as a legitimate company to recruit new members \cite{gemini}. At the same time, LockBit announced that they were looking for new affiliates through their leak site. In another bold move, LockBit tried to lure employees into sabotaging their organisations by providing them access to their networks in exchange for a share of the ransom, see Figure \ref{fig:lockbit_insiders} \cite{bleepingcomputerLockBitRansomware}. This approach has also been used by other ransomware groups, e.g., Lapsus\$ \cite{securityaffairsLapsusRansomware}, indicating that it could be successful.

Similarly, they may often resort to finding the proper legal representation to counter the group's legal issues, even across borders.   

Moreover, in ransomware groups, personnel are devoted to negotiatingng the ransom with the victims. While the negotiations might be in the form of short messages, these people have to collect information from accountants about the financial status of the victims and discuss with operators how to collect and use the proper decryptors. The latter stage is crucial, as part of the negotiations requires the decryption of a small set of files as proof that the purchased decryptor would be able to decrypt the files correctly. They also have to deal with operational issues that emerge with the transfer of files, errors with the decryptors and keys, victim assessment, and procedures to pay the ransom.

Finally, members of the group are in charge of public relations. Their goal is to be the voice of the group, develop their brand name, and handle relationships with other groups. However, they may also be assigned to make the proper connections with organisations, e.g., purchasing infrastructure, finding individuals to serve as insiders, or simply interacting with the authorities in the case of state-sponsored groups.

\subsection{Purchased capabilities}

\subsubsection{Marketplaces, forums, channels, and servers}
Illicit marketplaces and forums are usually hosted on the dark web. However, plenty of underground markets and forums are on the surface web \cite{lykousas2023cynicism}. Inside them, there are many entities of the MaaS ecosystem. Indeed, these marketplaces and forums are the primary sources for threat actors to recruit new members and seek new capabilities. Moreover, individuals and groups may use Telegram channels and Discord servers to advertise and sell their capabilities.

We start with malware authors who try to promote their coding skills by advertising the provision of their new malware. The malware may come as source code \cite{zdnetSourceCode,bleepingcomputerKnightRansomware} or builders \cite{darkreadingZeppelinRansomware} to further facilitate the threat actors. A builder is a software interface that allows the operator to pick and choose features to integrate into the malware, enabling customisation of the malware binary that will be produced from it. This way, even individuals with limited technical skills can launch sophisticated cyber attacks, democratising malware creation. 

Beyond malware authors, exploit developers seek to sell their exploits and may use similar methods to promote their capabilities. In this sense, they would sell exploits in the form of proof of concept that can allow privilege escalation, remote code execution, denial of service, etc. Here, the price varies depending on the impact the exploit can have, whether this is a known but unpatched vulnerability, but also the target audience in terms of size and financial status. For instance, a zero-day exploit for iPhones may cost far more than a zero-day exploit for a web server. 

Likewise, there are \emph{initial access brokers}, who act as middlemen in the malware supply chain. They are individuals or groups that specialise in obtaining and then selling unauthorised access to compromised systems, networks, or data to other criminals. To achieve this, they may use malware, phishing campaigns, or exploit vulnerabilities. Once access is obtained, initial access brokers advertise and sell this access to other threat actors. The price again varies significantly according to the perceived value of the target, the level of access obtained, and the potential for financial gain or espionage. Thus, initial access brokers significantly reduce the time to launch the malware on the target as they find possible targets, penetrate them, and delegate the control to their peers via remote desktop protocol (RDP), virtual private network (VPN), web shells, etc. By doing so, they enable malware deployment at scale.

The \emph{affiliates} are perhaps among the most crucial roles in the MaaS ecosystem. Practically, affiliates are individuals or groups that partner with threat groups and are responsible for spreading malware and infecting target systems. This can take multiple forms, from phishing campaigns and exploiting vulnerabilities in software and services to deploying malware on websites they penetrate or own. The relationship between affiliates and malware developers is typically based on a revenue-sharing model. When an affiliate successfully infects a system, and the victim pays a ransom (in the case of ransomware), or any financial gain is achieved through other forms of malware, the proceeds are split between the affiliate and the threat group. The specific split varies, but it is common for affiliates to receive a significant portion of the revenue, e.g., Himalaya ransomware advertised sharing 70\% of the ransom with their affiliates \cite{heimdalsecurityRansomwareGangs}. Essentially, this partnership has many implications; for instance, the threat group can drastically increase its capabilities without additional effort. Business-wise, outsourcing their hitmen allows them to monetise and achieve their goals and scale, targeting a wider range of victims across different sectors and geographical locations. Additionally, this model allows for specialisation within the MaaS ecosystem. For instance, malware authors can focus on improving their malware, while affiliates can specialise in finding innovative ways to breach security measures and infect systems. As a result, different affiliates may target different industries or deploy malware in unique ways, making it harder for defenders to anticipate and mitigate these threats. Finally, since it is the affiliates who distribute the malware, the threat actors can maintain a degree of separation from the actual illegal activities, potentially making it harder for law enforcement to trace the operation back to them or even block them.

Regarding malware distribution, traditional methods involve phishing, drive-by downloads, vulnerability exploitation, and hacking. Nevertheless, the affiliate program enables affiliates to distribute malware via compromised devices. The case of Triada malware \cite{googleblogFamilyHighlights,upstreamsystemsXHelperTriadaMalware} and other backdoored devices is a clear example of this practice \cite{bh2023,bleepingcomputerSecretBackdoor}. Affiliates may also use streaming services to deliver their malware \cite{lockett2023investigating} or compromise streaming apps and devices, which are not always very secure \cite{nikas2018know}. As a result, the affiliate concept paves the way for infiltration into the entire software and hardware supply chain, as compromised and corrupted individuals can be used to spread malware on an unprecedented scale.

\begin{figure}[th]
    \centering
    \includegraphics[width=\textwidth]{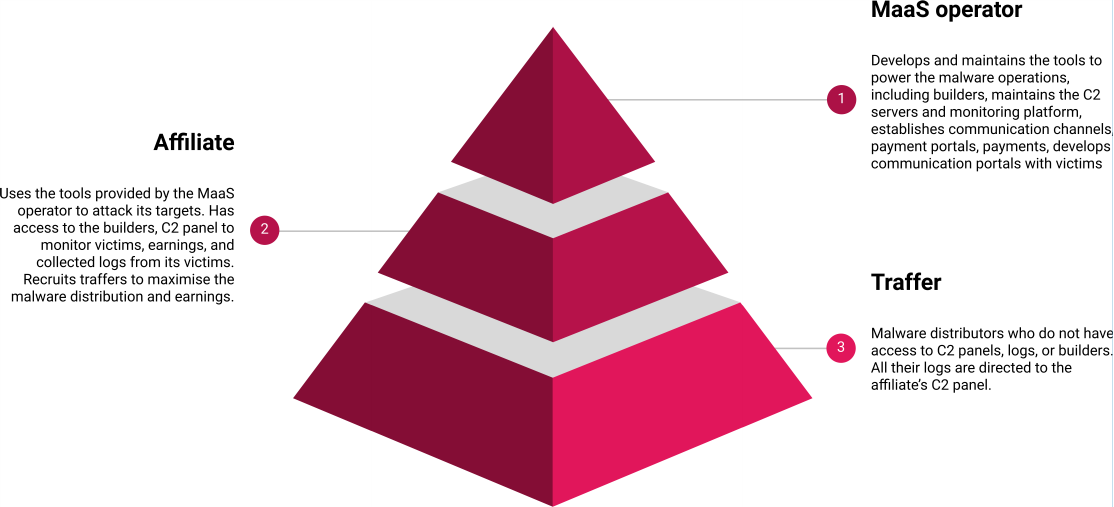}
    \caption{Hierarchy of the malware distribution in the MaaS ecosystem}
    \label{fig:hierarchy}
\end{figure}

\emph{Traffers} are cybercriminals recruited by affiliates to redirect internet traffic to sites that host the malware or to infect them themselves. The word comes from the Russian word \includegraphics[height=9pt]{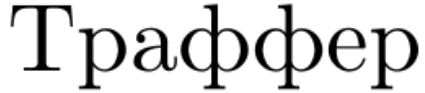} and refers to workers, more in a derogatory way. Since they are "workers", they are lower in the hierarchy; see Figure \ref{fig:hierarchy}, so they do not have many privileges in terms of, e.g., accessing the logs of the victims or the C2 panel of the Maas operator or botmaster. Most traffers seem to operate infostealers \cite{sekoiaTraffersDeep}, thus, they mainly harvest credentials, cards, and crypto wallets; however, they do not have access to them. As a result, the infections they cause can serve as an excellent first step to the organisation's infiltration process. Note that traffers must solely use the toolkit and malware provided by the affiliate, bring legitimate traffic (not bots), and try to stay below the radar of security providers. A special class of traffers is the log-checker who validate the credentials they stole. The profits of traffers stem from various sources, e.g., cryptocurrencies found in stolen wallets and selling the logs.  

\emph{Phishers} can have multiple roles in the MaaS ecosystem. Clearly, they can serve as initial access brokers by having compromised some victims and passing the control to a threat actor. Nevertheless, they can also serve as affiliates and simply distribute malware in their phishing campaigns. Furthermore, they can provide phishing kits to allow threat actors to masquerade as a legitimate organisation, either in terms of webpages or email-wise or use malvertising \cite{mandiantOpeningWhoop}. 

Similarly, \emph{penetration testers} assess whether specific hosts or infrastructures are susceptible to some vulnerability so that the affiliates can deliver the malware. Penetration testers may actively scan extensive networks or use services like Shodan to find vulnerable hosts or services. Once they verify their results, they pass their findings to affiliates, who may focus on compromising the network and its hosts to deliver the malware.

\emph{Carders} are groups or individuals specialised in the theft and fraudulent use of credit cards. In this sense, they serve as producers and consumers in the MaaS ecosystem. For instance, beyond phishing, hacking, and skimming devices at ATMs, they may use malware to steal this information from their victims, either in the form of keyloggers or as information stealers that collect the information from the browsers. Therefore, they use these means of communication to find the right malware authors to buy their products. On the other hand, they can be used by initial brokers for "card checking" services, that is, to verify the validity of stolen card details. Practically, each credit card has a different value depending on the card's limit, the issuing bank, and the cardholder's country. 

\begin{figure}[th]
    \centering
    \includegraphics[width=\textwidth]{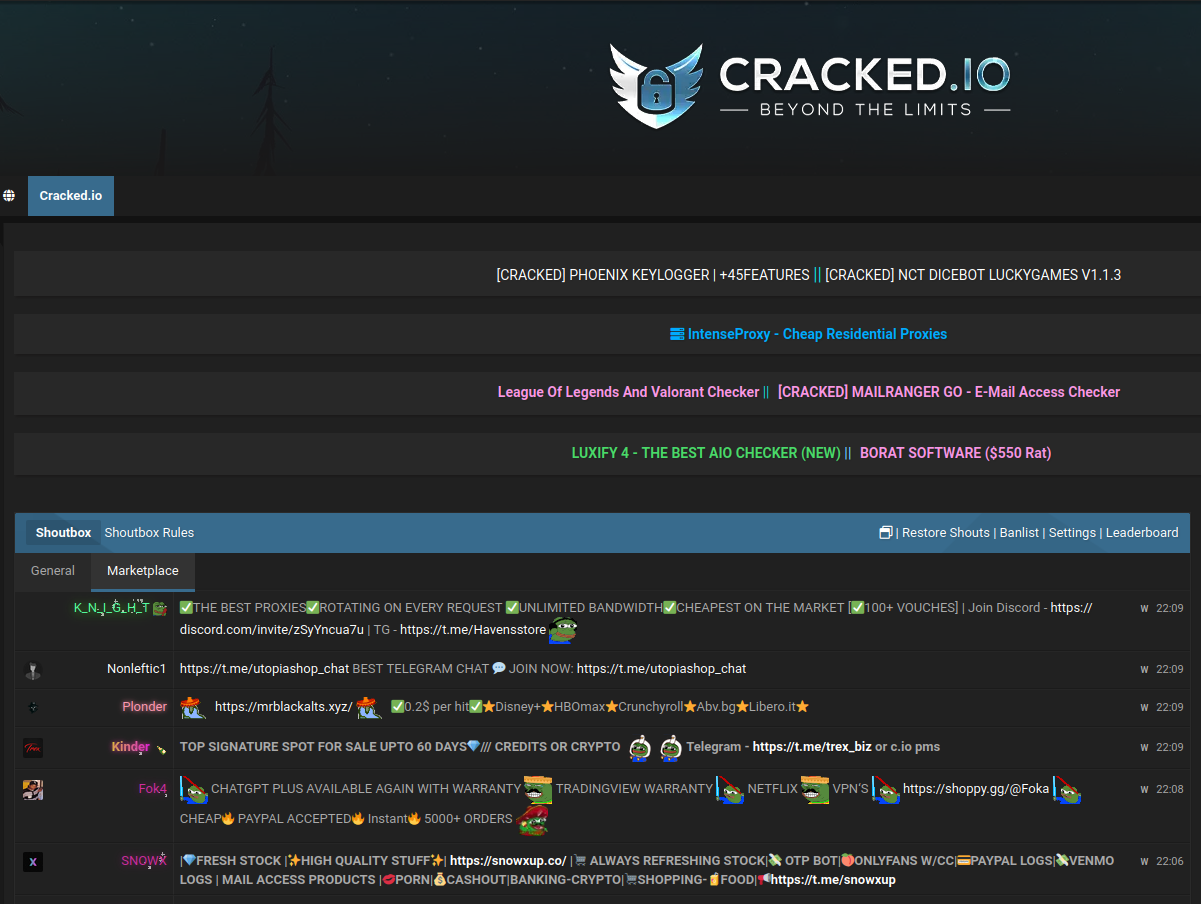}
    \caption{Promoting illegal services and selling compromised accounts on a hacking forum.}
    \label{fig:maas_ads}
\end{figure}

Finally, we have \emph{promoters}, which would advertise services, stolen cards and credentials, and malware on forums, marketplaces, forums, channels, and servers to foster revenue growth through marketing. They can be bots, bulletin boards, and individuals who try to promote their founder or services as traditional legitimate ones, see Figure \ref{fig:maas_ads}, and issue advertisements similar to those for legitimate products \cite{Cyble}.

\subsubsection{Infrastructure providers}

In terms of infrastructure providers, the MaaS ecosystem has a plethora of choices. Note that businesses may be used as service and infrastructure providers for malware distribution, management, control, and exfiltration as part of a malware campaign. This can be realised in various ways without the businesses being malicious, as this can be done without their knowledge or consent as part of the abuse of their services. Nevertheless, there are many occasions where companies simply turn a blind eye to cyber criminals and allow them to do their job \cite{Spamhaus,HALCYON}.

The most profound infrastructure providers are hosting service providers. In the context of MaaS, hosting service providers play a pivotal role in supporting various illegal activities online. These services, often referred to as "bulletproof hosting," provide the infrastructure necessary for cybercriminals to operate while actively avoiding compliance with law enforcement requests and ignoring abuse complaints. The hosting is needed to host malware, command and control (C2) servers and other malicious software. They are essential for managing botnets, distributing malware, and executing cyber attacks \cite{goncharov2015criminal,noroozian2019platforms,alrwais2017under}. Bulletproof hosting services are an upgrade to traditional hosting service providers as they are designed to provide anonymity to their clients. They often operate in jurisdictions with lax law enforcement or cybercrime laws, making it difficult for authorities to take legal action. These services use various techniques to evade detection and shutdown, including frequently moving their digital footprint across different countries. Indeed, they can quickly respond to shutdowns, domain seizures, and other enforcement actions by transferring their operations to new servers or locations, minimising downtime for cybercriminal operations. Their detection and identification are not always straightforward \cite{konte2015aswatch,alrwais2017under}, and several challenges emerge while taking them down.

\emph{Domain registrars} can play an unintentional yet significant role by providing the services needed for cybercriminals to register domain names used in their operations. While legitimate domain registrars operate within the bounds of the law and often have policies against misuse, the sheer volume of domains makes it challenging to prevent abuse entirely. Threat actors need domain names for various aspects of their operations, including hosting phishing sites, C2 servers for malware and their botnets, and fraudulent websites, to name a few. Note that many domain registrars offer privacy and proxy services that can hide the real identity of the domain registrant. While these services are legitimate and crucial for protecting individuals' privacy, they can also be abused by threat actors to obscure their identities and complicate law enforcement efforts \cite{casino2021unearthing}. Thus, they register domains through registrars that may have lax verification processes or that turn a blind eye to suspicious activities. The cases of malware using domain generation algorithms to resolve the C2 servers are particularly interesting. In such cases, the operators must register numerous domains, which in most cases are bogus, as they are randomly generated containing random patterns of numbers and letters \cite{nadji2013beheading,casino2021intercepting}, yet more advanced DGAs may use dictionaries to make them look less random \cite{10.5555/3241094.3241115,patsakis2021exploiting}. Similarly, threat actors may use various squatting methods to register domains that trick users or machines into visiting them instead of legitimate sites \cite{nikiforakis2013bitsquatting,agten2015seven,liu2018reexamination,quinkert2019s,kintis2017hiding}.

Cloud service providers (CSPs) have transformed the digital landscape by offering scalable, flexible, and efficient computing resources. Nevertheless, the features that make cloud services attractive for legitimate businesses also make them appealing for hosting malicious content on C2 servers. Moreover, cloud services can provide a layer of anonymity for threat actors, making tracing malicious activities back to their origins more challenging. Furthermore, the widespread trust in and reliance on cloud services can be exploited to bypass security measures; for example, network traffic to and from a reputable cloud provider might not be scrutinised as closely, allowing malicious traffic to blend in with legitimate traffic. For instance, ransomware groups abuse cloud service storage to upload exfiltrated data from compromised organisations. Similarly, in the context of DGAs, threat actors use cloud services to implement fast-flux networks, rapidly changing IP addresses and hostnames to evade detection and takedown efforts due to the agility and flexibility of CSPs. Similarly, threat actors exploit compromised cloud accounts to conduct their operations, taking advantage of the resources and trust associated with these accounts.

Botmasters play a central role in the MaaS ecosystem, managing networks of compromised computers (bots) to carry out a wide range of malicious activities. Once the botnet is established and managed through their C2 servers, they usually rent access to their botnet or even sell the whole of it to other threat actors, acting in this way as a service and infrastructure provider. They can offer their botnet for specific tasks, such as launching DDoS attacks, email spam campaigns, cryptocurrency mining, creating click farms, deploying other malware and getting a share of the earnings. A well-known case is the notorious Emotet, which establishes a botnet and then pushes various ransomware on compromised hosts \cite{patsakis2020analysing}. As a result, the rental and sale of botnet services create a profitable economic model within the MaaS ecosystem, allowing operators to monetise their malicious infrastructure. Even more, this also lowers the barrier to entry into cybercrime, as individuals can rent botnet services without needing the technical expertise to build and maintain a botnet by themselves. For more information on the monetisation mechanisms of botnets, the interested reader may refer to \cite{georgoulias2023market,georgoulias2023botnet}.

\begin{figure}[th]
    \centering
    \includegraphics[width=\textwidth]{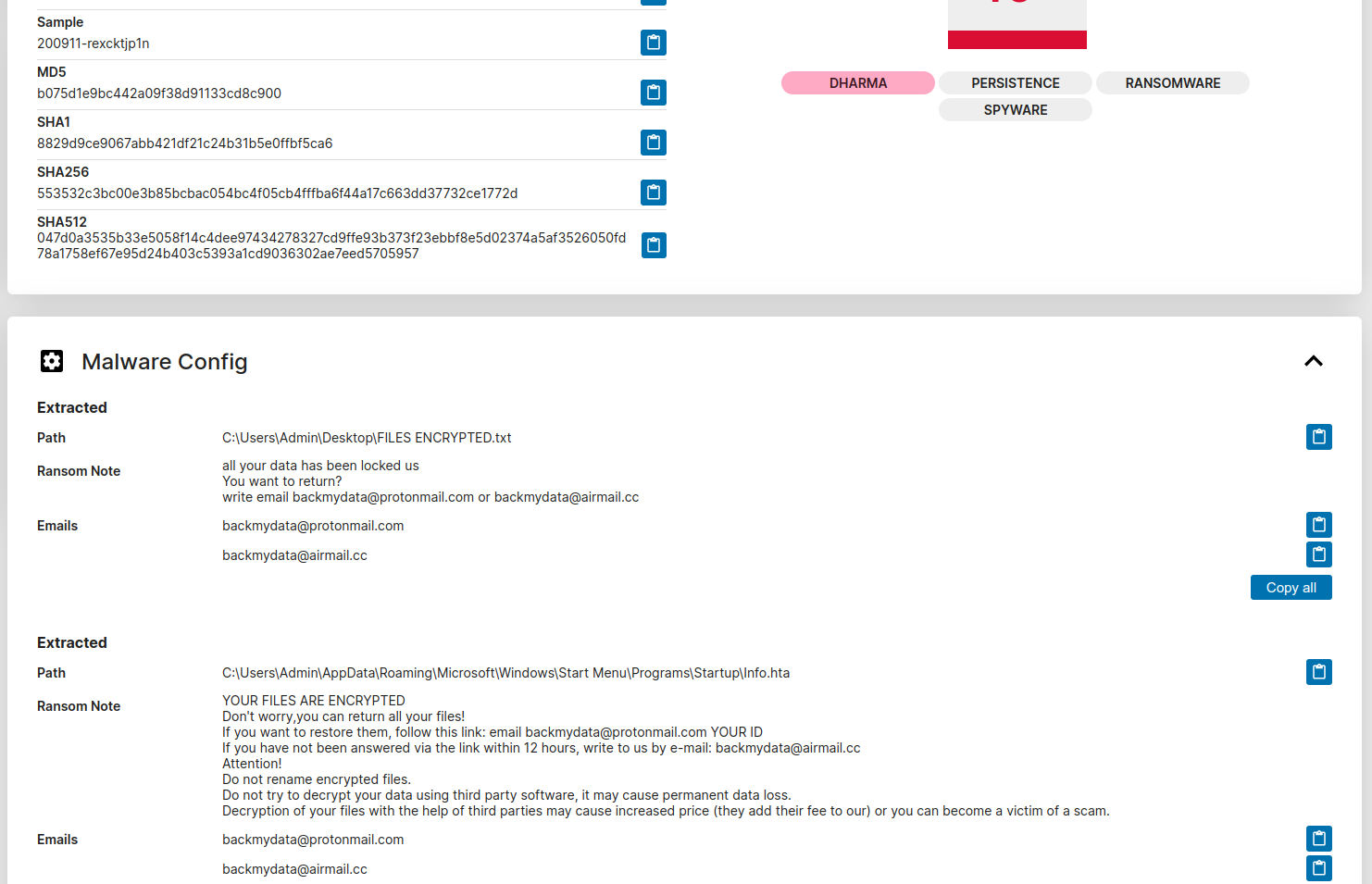}
    \caption{One of the first samples of the Dharma ransomware, using emails for the negotiations. Source: \protect\url{https://tria.ge/200911-rexcktjp1n}}
    \label{fig:dharma}
\end{figure}

Similarly to the previous ones, email providers, while primarily offering services for legitimate personal and business communication, can inadvertently play a role in the MaaS ecosystem. This involvement is mainly due to the misuse of email services by cybercriminals for various malicious activities. The case of phishing is the most obvious one, as email providers can be abused to deliver messages phishing emails with malicious attachments, typically malware droppers. Indeed, email remains a common vector for distributing malware. Moreover, business email compromise has become a huge threat as threat actors hack into or spoof company email accounts to impersonate executives, employees, or business partners to trick employees into transferring money to fraudulent accounts, extract confidential information, or convince them to perform malicious actions on their systems. Nevertheless, threat actors may use privacy-oriented email providers to communicate between them and with their victims. For instance, ransomware operators use such mail providers to negotiate the ransom with their victims in their initial steps in the "industry" when their "business" is not mature enough. In Figure \ref{fig:dharma}, we illustrate the analysis of one of the first samples of the Dharma ransomware (also known as CrySiS) back in 2020 from Hatching Triage. As it can be observed, the ransomware note has been extracted and invites victims to initiate discussions using such email providers. Notably, during the recent "Operation Cronos" to takedown Lockbit, 14,000 accounts on file-hosting service Mega and encrypted email providers Tutanota and Protonmail were shut down \cite{therecordLockBitTakedown}. The latter illustrates how much the MaaS ecosystem depends on such service and infrastructure providers. It should be pointed out the capability of MaaS operators such as Lockbit and Emotet to recover from international police operations and resume activity by identifying alternative service and infrastructure providers \cite{darkreadingEmotetRises,LockBitresurrect}.

VPN services are designed to enhance privacy and security online by encrypting users' internet traffic and masking their IP addresses. Evidently, the above is an ideal veil in the context of the MaaS ecosystem. First, VPNs can provide a layer of anonymity to threat actors by obscuring their real IP addresses, making it difficult for law enforcement and security researchers to trace malicious activities back to their sources. Moreover, this allows them to circumvent geo-restrictions and censorship, enabling access to blocked resources or markets for spreading malware, operating C2 servers, launching phishing campaigns, or distributing illegal content. In recent years, there has also been a rise in the exploitation of residential IP addresses as proxies \cite{tosun2021resip,10190651}, also exploiting backdoored devices \cite{bh2023}. This cross-border mixture and implication of residential proxies complicate jurisdictional and legal challenges for takedown and tracking efforts \cite{casino2022sok}.

Finally, there are dedicated infrastructures for malware authors to test their samples privately to minimise their detectability. Like VirusTotal, there are infrastructures to test malware samples with many antiviruses and collect responses. The key difference from VirusTotal is that these platforms do not disclose the results to others, especially the antimalware companies. This way, malware authors can fine-tune their executables without disclosing them, resting assure that their binaries will be undetected by many antimalware products.

\section{Conclusions}
In conclusion, the exploration of the MaaS ecosystem reveals a complex, multifaceted threat landscape that mirrors broader digital economic trends towards service orientation and commodification. This ecosystem facilitates a broad spectrum of cybercriminal activities, lowering the barrier to entry in terms of skills and infrastructure for cyber criminals and complicating the task of defence for individuals, organisations, and law enforcement authorities. The MaaS model, characterised by its efficiency, scalability, and adaptability, poses profound challenges to traditional cybersecurity measures. It requires a paradigm shift towards more dynamic, behaviour-based, and AI-enhanced defence strategies. For instance, in the past, threat groups had more distinctive footprints when using specific TTPs and tooling. Nevertheless, the use of arbitrary affiliates blurs this approach as affiliates may use radically different methods, tooling, and tradecraft, so requiring optimised detection and adaptive capabilities that span across all attack stages. This scenario is even more challenging if we consider the complexities associated with the proliferation of mobile devices, mobile edge computing, third-party AI tools and resources, and the increasing adoption of decentralized finance solutions.  

Our analysis underscores the critical role of various actors within the MaaS ecosystem, including malware developers, affiliates, initial access brokers, and infrastructure providers, all of whom contribute to the proliferation of cyber threats, directly or indirectly. Furthermore, the research points to the importance of understanding the economic and motivational nuances within the evolving MaaS ecosystem. This understanding is crucial for developing targeted strategies that disrupt the financial incentives that drive the proliferation of MaaS offerings.

Considering MaaS as an investment that criminals would pursue, it is highly profitable as it has an extremely low cost of goods and maintenance but a high operating margin. The above, coupled with the fact that the prosecution can be delayed or avoided due to the complications of cross-jurisdiction legal issues and the blurring of digital evidence through various security and privacy measures, justifies the huge surge of this underground industry. Therefore, enhanced international cooperation is necessary, bringing forward the need for more robust and unified legal frameworks and developing advanced technological solutions to detect, mitigate, and prevent the threats posed by this evolving cybercrime model.

The above demonstrates that the fight against MaaS is not straightforward or a matter of a single domain. The problem is not simply improving the malware detection mechanisms; as there is a dynamic and vibrant ecosystem. Dismantling MaaS requires a multidisciplinary approach, considering the cooperation of technological, regulatory, legal, and financial stakeholders to devise more effective countermeasures against the scourge of commoditised malware in all these domains. Furthermore, education is critical to a proper cyber-resilience strategy against. In other words, MaaS cannot be effectively contained without solid cyber awareness and cyber hygiene programs for the whole society.

\bibliographystyle{plain}
\bibliography{refs}

\begin{thebibliography}{10}

\bibitem{bleepingcomputerLockBitRansomware}
Lawrence Abrams.
\newblock {L}ock{B}it ransomware recruiting insiders to breach corporate networks.
\newblock \url{https://www.bleepingcomputer.com/news/security/lockbit-ransomware-recruiting-insiders-to-breach-corporate-networks/}, 2021.
\newblock [Accessed 05-03-2024].

\bibitem{trendmicroRansomwareDouble}
Janus Agcaoili, Miguel Ang, Earle Earnshaw, Byron Gelera, and Nikko Tamaña.
\newblock Ransomware double extortion and beyond: {R}{E}vil, {C}lop, and {C}onti - {S}ecurity {N}ews --- trendmicro.com.
\newblock \url{https://www.trendmicro.com/vinfo/us/security/news/cybercrime-and-digital-threats/ransomware-double-extortion-and-beyond-revil-clop-and-conti}, 2021.
\newblock [Accessed 05-03-2024].

\bibitem{agten2015seven}
Pieter Agten, Wouter Joosen, Frank Piessens, and Nick Nikiforakis.
\newblock Seven months' worth of mistakes: A longitudinal study of typosquatting abuse.
\newblock In {\em Proceedings of the 22nd Network and Distributed System Security Symposium (NDSS 2015)}. Internet Society, 2015.

\bibitem{alrwais2017under}
Sumayah Alrwais, Xiaojing Liao, Xianghang Mi, Peng Wang, XiaoFeng Wang, Feng Qian, Raheem Beyah, and Damon McCoy.
\newblock Under the shadow of sunshine: Understanding and detecting bulletproof hosting on legitimate service provider networks.
\newblock In {\em 2017 IEEE Symposium on Security and Privacy (SP)}, pages 805--823. IEEE, 2017.

\bibitem{ANDRADE2019102352}
Roberto~O Andrade and Sang~Guun Yoo.
\newblock Cognitive security: A comprehensive study of cognitive science in cybersecurity.
\newblock {\em Journal of Information Security and Applications}, 48:102352, 2019.

\bibitem{talosintelligenceLazarusTale}
Jungsoo~An Asheer~Malhotra, Vitor~Ventura.
\newblock {L}azarus and the tale of three {R}{A}{T}s.
\newblock \url{https://blog.talosintelligence.com/lazarus-three-rats/}, 2022.
\newblock [Accessed 06-03-2024].

\bibitem{bethany2024large}
Mazal Bethany, Athanasios Galiopoulos, Emet Bethany, Mohammad~Bahrami Karkevandi, Nishant Vishwamitra, and Peyman Najafirad.
\newblock Large language model lateral spear phishing: A comparative study in large-scale organizational settings.
\newblock {\em arXiv preprint arXiv:2401.09727}, 2024.

\bibitem{lumenEmotetIlluminated}
{Black Lotus Labs}.
\newblock {E}motet illuminated: {M}apping a tiered botnet using global network forensics.
\newblock \url{https://blog.lumen.com/emotet-illuminated-mapping-a-tiered-botnet-using-global-network-forensics/}, 2019.
\newblock [Accessed 04-03-2024].

\bibitem{ransomwheRansomwhere}
Jack Cable.
\newblock {R}ansomwhere --- ransomwhe.re.
\newblock \url{https://ransomwhe.re/}.
\newblock [Accessed 04-03-2024].

\bibitem{carpentier2023mapping}
Catherine Carpentier-Desjardins, Masarah Paquet-Clouston, Stefan Kitzler, and Bernhard Haslhofer.
\newblock Mapping the defi crime landscape: An evidence-based picture.
\newblock {\em arXiv preprint arXiv:2310.04356}, 2023.

\bibitem{casino2021intercepting}
Fran Casino, Nikolaos Lykousas, Ivan Homoliak, Constantinos Patsakis, and Julio Hernandez-Castro.
\newblock Intercepting hail hydra: Real-time detection of algorithmically generated domains.
\newblock {\em Journal of Network and Computer Applications}, 190:103135, 2021.

\bibitem{casino2021unearthing}
Fran Casino, Nikolaos Lykousas, Vasilios Katos, and Constantinos Patsakis.
\newblock Unearthing malicious campaigns and actors from the blockchain dns ecosystem.
\newblock {\em Computer Communications}, 179:217--230, 2021.

\bibitem{casino2022sok}
Fran Casino, Claudia Pina, Pablo L{\'o}pez-Aguilar, Edgar Batista, Agusti Solanas, and Constantinos Patsakis.
\newblock Sok: cross-border criminal investigations and digital evidence.
\newblock {\em Journal of Cybersecurity}, 8(1):tyac014, 2022.

\bibitem{chainreport}
Chainalysis.
\newblock The chainalysis 2023 crypto crime report.
\newblock \url{https://go.chainalysis.com/2023-crypto-crime-report.html}, 2023.

\bibitem{bb}
{Check Point}.
\newblock {Check Point Research exposes new versions of the BBTok banking malware, which targets clients of over 40 Mexican and Brazilian banks}.
\newblock \url{https://blog.checkpoint.com/security/check-point-research-exposes-new-versions-of-the-bbtok-banking-malware-which-targets-clients-of-over-40-mexican-and-brazilian-banks/}, 2023.
\newblock [Accessed 08-03-2024].

\bibitem{10190651}
Elisa Chiapponi, Marc Dacier, and Olivier Thonnard.
\newblock Inside residential {IP} proxies: Lessons learned from large measurement campaigns.
\newblock In {\em {IEEE} European Symposium on Security and Privacy, EuroS{\&}P 2023 - Workshops, Delft, Netherlands, July 3-7, 2023}, pages 501--512. {IEEE}, 2023.

\bibitem{bleepingcomputerSecretBackdoor}
Catalin Cimpanu.
\newblock Secret backdoor in some low-priced {Android} phones sent data to a server in {China}.
\newblock \url{https://www.bleepingcomputer.com/news/security/secret-backdoor-in-some-low-priced-android-phones-sent-data-to-a-server-in-china/}.
\newblock [Accessed 04-03-2024].

\bibitem{zdnetSourceCode}
Catalin Cimpanu.
\newblock {S}ource code of {D}harma ransomware pops up for sale on hacking forums.
\newblock \url{https://www.zdnet.com/article/source-code-of-dharma-ransomware-pops-up-for-sale-on-hacking-forums/}, 2020.
\newblock [Accessed 04-03-2024].

\bibitem{10.1145/3585385}
Antonio~Emanuele Cin\`{a}, Kathrin Grosse, Ambra Demontis, Sebastiano Vascon, Werner Zellinger, Bernhard~A. Moser, Alina Oprea, Battista Biggio, Marcello Pelillo, and Fabio Roli.
\newblock Wild patterns reloaded: A survey of machine learning security against training data poisoning.
\newblock {\em ACM Comput. Surv.}, 55(13s), jul 2023.

\bibitem{certCERTUA2}
{Computer Emergency Response Team of Ukraine}.
\newblock {CERT-UA\#7469)}.
\newblock \url{https://cert.gov.ua/article/5702579}, 2023.
\newblock [Accessed 06-03-2024].

\bibitem{certCERTUA}
{Computer Emergency Response Team of Ukraine}.
\newblock {CERT-UA\#7627)}.
\newblock \url{https://cert.gov.ua/article/6123309}, 2023.
\newblock [Accessed 06-03-2024].

\bibitem{Cyble}
Cyble.
\newblock {D}uck{L}ogs – {N}ew malware strain spotted in the wild.
\newblock \url{https://cyble.com/blog/ducklogs-new-malware-strain-spotted-in-the-wild/}, 2022.
\newblock [Accessed 05-03-2024].

\bibitem{elliptic}
Elliptic.
\newblock Financial crime typologies in cryptoassets.
\newblock \url{https://www.elliptic.co/resources/typologies-concise-guide-crypto-leaders}, 2020.

\bibitem{europaDarkMarketWorlds}
Europol.
\newblock {D}ark{M}arket: world's largest illegal dark web marketplace taken down.
\newblock \url{https://www.europol.europa.eu/media-press/newsroom/news/darkmarket-worlds-largest-illegal-dark-web-marketplace-taken-down}, 2021.
\newblock [Accessed 04-03-2024].

\bibitem{europaDarkVendors}
Europol.
\newblock 288 dark web vendors arrested in major marketplace seizure.
\newblock \url{https://www.europol.europa.eu/media-press/newsroom/news/288-dark-web-vendors-arrested-in-major-marketplace-seizure}, 2023.
\newblock [Accessed 04-03-2024].

\bibitem{europaTakedownNotorious}
Europol.
\newblock Takedown of notorious hacker marketplace selling your identity to criminals.
\newblock \url{https://www.europol.europa.eu/media-press/newsroom/news/takedown-of-notorious-hacker-marketplace-selling-your-identity-to-criminals}, 2023.
\newblock [Accessed 04-03-2024].

\bibitem{generative_apts}
Fortune.
\newblock Microsoft says iran, north korea, russia and china are beginning to use generative ai in offensive cyberattacks, 2024.

\bibitem{cybersecurityventuresCybercrimeCost}
Di~Freeze.
\newblock {C}ybercrime to cost the world 8 trillion annually in 2023.
\newblock \url{https://cybersecurityventures.com/cybercrime-to-cost-the-world-8-trillion-annually-in-2023/}, 2022.
\newblock [Accessed 04-03-2024].

\bibitem{gemini}
{Gemini Advisory}.
\newblock {FIN}7 recruits talent for push into ransomware.
\newblock \url{https://geminiadvisory.io/fin7-ransomware-bastion-secure/}, 2021.
\newblock [Accessed 05-03-2024].

\bibitem{georgoulias2023botnet}
Dimitrios Georgoulias, Jens~Myrup Pedersen, Morten Falch, and Emmanouil Vasilomanolakis.
\newblock Botnet business models, takedown attempts, and the darkweb market: a survey.
\newblock {\em ACM Computing Surveys}, 55(11):1--39, 2023.

\bibitem{georgoulias2023market}
Dimitrios Georgoulias, Jens~Myrup Pedersen, Alice Hutchings, Morten Falch, and Emmanouil Vasilomanolakis.
\newblock In the market for a botnet? an in-depth analysis of botnet-related listings on darkweb marketplaces.
\newblock In {\em Symposium on Electronic Crime Research 2023}, 2023.

\bibitem{goncharov2015criminal}
Max Goncharov.
\newblock Criminal hideouts for lease: Bulletproof hosting services.
\newblock {\em Forward-Looking Threat Research (FTR) Team, A TrendLabsSM Research Paper}, 28, 2015.

\bibitem{googleblogFamilyHighlights}
{Google Security Blog}.
\newblock {P}{H}{A} {F}amily {H}ighlights: {T}riada.
\newblock \url{https://security.googleblog.com/2019/06/pha-family-highlights-triada.html}, 2019.
\newblock [Accessed 04-03-2024].

\bibitem{godf}
Artem Grischenko.
\newblock Godfather: A banking trojan that is impossible to refuse.
\newblock \url{https://www.group-ib.com/blog/godfather-trojan/}, 2022.
\newblock [Accessed 08-03-2024].

\bibitem{guri2022air}
Mordechai Guri.
\newblock Air-fi: Leaking data from air-gapped computers using wi-fi frequencies.
\newblock {\em IEEE Transactions on Dependable and Secure Computing}, 2022.

\bibitem{guri2019air}
Mordechai Guri and Dima Bykhovsky.
\newblock air-jumper: Covert air-gap exfiltration/infiltration via security cameras \& infrared (ir).
\newblock {\em Computers \& Security}, 82:15--29, 2019.

\bibitem{guri2018bridgeware}
Mordechai Guri and Yuval Elovici.
\newblock Bridgeware: The air-gap malware.
\newblock {\em Communications of the ACM}, 61(4):74--82, 2018.

\bibitem{GURI2020101721}
Mordechai Guri, Yosef Solewicz, and Yuval Elovici.
\newblock Fansmitter: Acoustic data exfiltration from air-gapped computers via fans noise.
\newblock {\em Computers \& Security}, 91:101721, 2020.

\bibitem{guri2019powerhammer}
Mordechai Guri, Boris Zadov, Dima Bykhovsky, and Yuval Elovici.
\newblock Powerhammer: Exfiltrating data from air-gapped computers through power lines.
\newblock {\em IEEE Transactions on Information Forensics and Security}, 15:1879--1890, 2019.

\bibitem{HALCYON}
Halcyon.
\newblock Cloudzy with a chance of ransomware, 2023.

\bibitem{Haslhofer2021a}
Bernhard Haslhofer, Rainer Stütz, Matteo Romiti, and Ross King.
\newblock Graphsense: A general-purpose cryptoasset analytics platform.
\newblock {\em Arxiv pre-print}, 2021.

\bibitem{heimdalsecurityRansomwareGangs}
Heimdal.
\newblock {R}ansomware gangs are now using new recruitment strategies.
\newblock \url{https://heimdalsecurity.com/blog/ransomware-gangs-recruitment-strategies/}, 2021.
\newblock [Accessed 05-03-2024].

\bibitem{ho2019detecting}
Grant Ho, Asaf Cidon, Lior Gavish, Marco Schweighauser, Vern Paxson, Stefan Savage, Geoffrey~M Voelker, and David Wagner.
\newblock Detecting and characterizing lateral phishing at scale.
\newblock In {\em 28th USENIX Security Symposium (USENIX Security 19)}, pages 1273--1290, 2019.

\bibitem{10.1145/3199674}
Keman Huang, Michael Siegel, and Stuart Madnick.
\newblock Systematically understanding the cyber attack business: A survey.
\newblock {\em ACM Comput. Surv.}, 51(4), jul 2018.

\bibitem{joint2011sp}
Joint Task Force~Transformation Initiative et~al.
\newblock {\em SP 800-39. managing information security risk: Organization, mission, and information system view}.
\newblock National Institute of Standards \& Technology, 2011.

\bibitem{karapapas2020ransomware}
Christos Karapapas, Iakovos Pittaras, Nikos Fotiou, and George~C Polyzos.
\newblock Ransomware as a service using smart contracts and ipfs.
\newblock In {\em 2020 IEEE International Conference on Blockchain and Cryptocurrency (ICBC)}, pages 1--5. IEEE, 2020.

\bibitem{kasper}
Kaspersky.
\newblock {Understanding Malware-as-a-Service}.
\newblock \url{https://securelist.com/malware-as-a-service-market/109980/}, 2022.
\newblock [Accessed 05-03-2024].

\bibitem{kintis2017hiding}
Panagiotis Kintis, Najmeh Miramirkhani, Charles Lever, Yizheng Chen, Rosa Romero-G{\'o}mez, Nikolaos Pitropakis, Nick Nikiforakis, and Manos Antonakakis.
\newblock Hiding in plain sight: A longitudinal study of combosquatting abuse.
\newblock In {\em Proceedings of the 2017 ACM SIGSAC Conference on Computer and Communications Security}, pages 569--586, 2017.

\bibitem{konte2015aswatch}
Maria Konte, Roberto Perdisci, and Nick Feamster.
\newblock Aswatch: An as reputation system to expose bulletproof hosting ases.
\newblock In {\em Proceedings of the 2015 ACM Conference on Special Interest Group on Data Communication}, pages 625--638, 2015.

\bibitem{koutsokostas2022invoice}
Vasilios Koutsokostas, Nikolaos Lykousas, Theodoros Apostolopoulos, Gabriele Orazi, Amrita Ghosal, Fran Casino, Mauro Conti, and Constantinos Patsakis.
\newblock Invoice\# 31415 attached: Automated analysis of malicious microsoft office documents.
\newblock {\em Computers \& Security}, 114:102582, 2022.

\bibitem{langner2011stuxnet}
Ralph Langner.
\newblock Stuxnet: Dissecting a cyberwarfare weapon.
\newblock {\em IEEE Security \& Privacy}, 9(3):49--51, 2011.

\bibitem{cccOperationTriangulation}
Boris Larin, Leonid Bezvershenko, and Georgy Kucherin.
\newblock {O}peration {T}riangulation: What you get when attack i{P}hones of researchers.
\newblock \url{https://media.ccc.de/v/37c3-11859-operation_triangulation_what_you_get_when_attack_iphones_of_researchers}, 2023.
\newblock [Accessed 08-03-2024].

\bibitem{darkreadingEmotetRises}
Robert Lemos.
\newblock {E}motet rises again with more aophistication, evasion.
\newblock \url{https://www.darkreading.com/threat-intelligence/emotet-rises-again-with-more-sophistication-evasion}, 2022.
\newblock [Accessed 08-03-2024].

\bibitem{liu2018reexamination}
Baojun Liu, Chaoyi Lu, Zhou Li, Ying Liu, Hai-Xin Duan, Shuang Hao, and Zaifeng Zhang.
\newblock A reexamination of internationalized domain names: The good, the bad and the ugly.
\newblock In {\em DSN}, pages 654--665, 2018.

\bibitem{lockett2023investigating}
Adam Lockett, Ioannis Chalkias, Cagatay Yucel, Jane Henriksen-Bulmer, and Vasilis Katos.
\newblock Investigating iptv malware in the wild.
\newblock {\em Future Internet}, 15(10):325, 2023.

\bibitem{lykousas2023cynicism}
Nikolaos Lykousas, Vasilios Koutsokostas, Fran Casino, and Constantinos Patsakis.
\newblock The cynicism of modern cybercrime: Automating the analysis of surface web marketplaces.
\newblock In {\em 2023 IEEE International Conference on Service-Oriented System Engineering (SOSE)}, pages 161--171. IEEE, 2023.

\bibitem{Mandiant}
MANDIANT.
\newblock Unc2452{ merged into APT29: Russia-based Espionage Group}, 2023.

\bibitem{therecordLockBitTakedown}
Alexander Martin.
\newblock {L}ock{B}it takedown: {P}olice shut more than 14,000 accounts on {M}ega, {T}utanota and {P}rotonmail.
\newblock \url{https://therecord.media/lockbit-ransomware-takedown-mega-tutanota-protonmail}, 2024.
\newblock [Accessed 05-03-2024].

\bibitem{meland2020ransomware}
Per~H{\aa}kon Meland, Yara Fareed~Fahmy Bayoumy, and Guttorm Sindre.
\newblock The ransomware-as-a-service economy within the darknet.
\newblock {\em Computers \& Security}, 92:101762, 2020.

\bibitem{microsoftDefineRansomware}
{Microsoft Security}.
\newblock Define ransomware, human-operated ransomware, and how to prevent ransomware cyber attack.
\newblock \url{https://learn.microsoft.com/en-us/security/ransomware/human-operated-ransomware}, 2024.
\newblock [Accessed 04-03-2024].

\bibitem{nadji2013beheading}
Yacin Nadji, Manos Antonakakis, Roberto Perdisci, David Dagon, and Wenke Lee.
\newblock Beheading hydras: performing effective botnet takedowns.
\newblock In {\em Proceedings of the 2013 ACM SIGSAC conference on Computer \& communications security}, pages 121--132, 2013.

\bibitem{nikas2018know}
Alexios Nikas, Efthimios Alepis, and Constantinos Patsakis.
\newblock I know what you streamed last night: On the security and privacy of streaming.
\newblock {\em Digital Investigation}, 25:78--89, 2018.

\bibitem{nikiforakis2013bitsquatting}
Nick Nikiforakis, Steven Van~Acker, Wannes Meert, Lieven Desmet, Frank Piessens, and Wouter Joosen.
\newblock Bitsquatting: Exploiting bit-flips for fun, or profit?
\newblock In {\em Proceedings of the 22nd international conference on World Wide Web}, pages 989--998, 2013.

\bibitem{cybereasonPowerLessTrojan}
Cybereason Nocturnus.
\newblock {P}ower{L}ess trojan: {I}ranian {A}{P}{T} {P}hosphorus adds new {P}ower{S}hell backdoor for espionage.
\newblock \url{https://www.cybereason.com/blog/research/powerless-trojan-iranian-apt-phosphorus-adds-new-powershell-backdoor-for-espionage}, 2022.
\newblock [Accessed 06-03-2024].

\bibitem{noroozian2019platforms}
Arman Noroozian, Jan Koenders, Eelco Van~Veldhuizen, Carlos~H Ganan, Sumayah Alrwais, Damon McCoy, and Michel Van~Eeten.
\newblock Platforms in everything: Analyzing $\{$Ground-Truth$\}$ data on the anatomy and economics of $\{$Bullet-Proof$\}$ hosting.
\newblock In {\em 28th USENIX Security Symposium (USENIX Security 19)}, pages 1341--1356, 2019.

\bibitem{CACM-Ransomware}
Kris Oosthoek, Jack Cable, and Georgios Smaragdakis.
\newblock {A Tale of Two Markets: Investigating the Ransomware Payments Economy}.
\newblock {\em {Communications of the ACM}}, 66(8), 2023.

\bibitem{10012354}
Kris Oosthoek, Mark Van~Staalduinen, and Georgios Smaragdakis.
\newblock Quantifying dark web shops’ illicit revenue.
\newblock {\em IEEE Access}, 11:4794--4808, 2023.

\bibitem{securityaffairsLapsusRansomware}
Pierluigi Paganini.
\newblock {L}apsus\$ ransomware group announced recruitment of insiders.
\newblock \url{https://securityaffairs.com/128912/cyber-crime/lapsus-ransomware-is-hiring.html}, 2022.
\newblock [Accessed 05-03-2024].

\bibitem{patsakis2021exploiting}
Constantinos Patsakis and Fran Casino.
\newblock Exploiting statistical and structural features for the detection of domain generation algorithms.
\newblock {\em Journal of Information Security and Applications}, 58:102725, 2021.

\bibitem{patsakis2020analysing}
Constantinos Patsakis and Anargyros Chrysanthou.
\newblock Analysing the fall 2020 emotet campaign.
\newblock {\em arXiv preprint arXiv:2011.06479}, 2020.

\bibitem{patsakis2023cashing}
Constantinos Patsakis, Eugenia Politou, Efthimios Alepis, and Julio Hernandez-Castro.
\newblock Cashing out crypto: state of practice in ransom payments.
\newblock {\em International Journal of Information Security}, pages 1--14, 2023.

\bibitem{10.5555/3241094.3241115}
Daniel Plohmann, Khaled Yakdan, Michael Klatt, Johannes Bader, and Elmar Gerhards-Padilla.
\newblock A comprehensive measurement study of domain generating malware.
\newblock In {\em Proceedings of the 25th USENIX Conference on Security Symposium}, SEC'16, page 263–278, USA, 2016. USENIX Association.

\bibitem{sekoiaTraffersDeep}
Livia~Tibirna Quentin~Bourgue and TDR (Threat Detection~\& Research).
\newblock {T}raffers: a deep dive into the information stealer ecosystem --- blog.sekoia.io.
\newblock \url{https://blog.sekoia.io/traffers-a-deep-dive-into-the-information-stealer-ecosystem/}, 2022.
\newblock [Accessed 05-03-2024].

\bibitem{quinkert2019s}
Florian Quinkert, Tobias Lauinger, William Robertson, Engin Kirda, and Thorsten Holz.
\newblock It's not what it looks like: Measuring attacks and defensive registrations of homograph domains.
\newblock In {\em 2019 IEEE Conference on Communications and Network Security (CNS)}, pages 259--267. IEEE, 2019.

\bibitem{reid2012analysis}
Fergal Reid and Martin Harrigan.
\newblock An analysis of anonymity in the bitcoin system.
\newblock {\em Security and Privacy in Social Networks}, page 197, 2012.

\bibitem{ron2013quantitative}
Dorit Ron and Adi Shamir.
\newblock Quantitative analysis of the full bitcoin transaction graph.
\newblock In {\em Financial Cryptography and Data Security: 17th International Conference, FC 2013, Okinawa, Japan, April 1-5, 2013, Revised Selected Papers 17}, pages 6--24. Springer, 2013.

\bibitem{LockBitresurrect}
Stefanie Schappert.
\newblock {L}ock{B}it back online, already targeting hospitals with ransomware.
\newblock \url{https://cybernews.com/news/lockbit-back-online-already-targeting-hospitals-with-ransomware/}, 2024.
\newblock [Accessed 08-03-2024].

\bibitem{skopik2020under}
Florian Skopik and Timea Pahi.
\newblock Under false flag: using technical artifacts for cyber attack attribution.
\newblock {\em Cybersecurity}, 3:1--20, 2020.

\bibitem{Spamhaus}
Spamhaus.
\newblock Russian registrar {N}{A}{U}{N}{E}{T} knowingly harbours cybercriminals.
\newblock \url{https://www.spamhaus.org/resource-hub/service-providers/russian-registrar-naunet-knowingly-harbours-cybercriminals/}, 2012.
\newblock [Accessed 06-03-2024].

\bibitem{teamcymruTrackingBokBot}
{Team Cymru}.
\newblock Tracking bokbot ({I}ced{I}{D}) infrastructure.
\newblock \url{https://www.team-cymru.com/post/tracking-bokbot-icedid-infrastructuremapping-a-vast-and-currently-active-icedid-network}, 2021.
\newblock [Accessed 04-03-2024].

\bibitem{thehackernewsQakBotMalware}
{The hacker news}.
\newblock {Q}ak{B}ot malware operators expand c2 network with 15 new servers.
\newblock \url{https://thehackernews.com/2023/08/qakbot-malware-operators-expand-c2.html}, 2023.
\newblock [Accessed 04-03-2024].

\bibitem{thornton2019politics}
Ian Thornton-Trump.
\newblock The politics of cyber.
\newblock {\em EDPACS}, 59(3):1--17, 2019.

\bibitem{chameleon}
{ThreatFabric}.
\newblock Android banking trojan chameleon can now bypass any biometric authentication.
\newblock \url{https://www.threatfabric.com/blogs/android-banking-trojan-chameleon-is-back-in-action}, 2023.
\newblock [Accessed 08-03-2024].

\bibitem{mandiantOpeningWhoop}
Ryan Tomcik, Adrian McCabe, Rufus Brown, and Geoff Ackerman.
\newblock Opening a can of whoop ads: Detecting and disrupting a malvertising campaign distributing backdoors.
\newblock \url{https://www.mandiant.com/resources/blog/detecting-disrupting-malvertising-backdoors}, 2023.
\newblock [Accessed 05-03-2024].

\bibitem{tosun2021resip}
Altug Tosun, Michele De~Donno, Nicola Dragoni, and Xenofon Fafoutis.
\newblock Resip host detection: identification of malicious residential ip proxy flows.
\newblock In {\em 2021 IEEE International Conference on Consumer Electronics (ICCE)}, pages 1--6. IEEE, 2021.

\bibitem{bleepingcomputerKnightRansomware}
Bill Toulas.
\newblock {K}night ransomware source code for sale after leak site shuts down.
\newblock \url{https://www.bleepingcomputer.com/news/security/knight-ransomware-source-code-for-sale-after-leak-site-shuts-down/}, 2024.
\newblock [Accessed 04-03-2024].

\bibitem{upstreamsystemsXHelperTriadaMalware}
upstream.
\newblock x{H}elper/{T}riada malware pre-installed on thousands of low cost {C}hinese {A}ndroid devices in emerging markets - {U}pstream.
\newblock \url{https://www.upstreamsystems.com/press/press-releases/xhelper-triada-malware-pre-installed-on-thousands-of-low-cost-chinese-android-devices-in-emerging-markets/}, 2020.
\newblock [Accessed 04-03-2024].

\bibitem{justiceJusticeDepartment}
{U.S. Department of Justice}.
\newblock Justice department investigation leads to shutdown of largest online darknet marketplace.
\newblock \url{https://www.justice.gov/opa/pr/justice-department-investigation-leads-shutdown-largest-online-darknet-marketplace}, 2022.
\newblock [Accessed 04-03-2024].

\bibitem{darkreadingZeppelinRansomware}
Jai Vijayan.
\newblock {Z}eppelin ransomware source code \& builder sells for \$500 on dark web.
\newblock \url{https://www.darkreading.com/ics-ot-security/zeppelin-ransomware-source-code-builder-sells-500-dark-web}, 2024.
\newblock [Accessed 04-03-2024].

\bibitem{bh2023}
Fyodor Yarochkin, Zhengyu Dong, Vladimir Kropotov, and Paul Pajares.
\newblock Behind the scenes: How criminal enterprises pre-infect millions of mobile devices.
\newblock In {\em BlackHat ASIA}, 2023.

\bibitem{9581257}
Xinyang Zhang, Zheng Zhang, Shouling Ji, and Ting Wang.
\newblock Trojaning language models for fun and profit.
\newblock In {\em 2021 IEEE European Symposium on Security and Privacy (EuroS\&P)}, pages 179--197, 2021.

\bibitem{zhou2023sok}
Liyi Zhou, Xihan Xiong, Jens Ernstberger, Stefanos Chaliasos, Zhipeng Wang, Ye~Wang, Kaihua Qin, Roger Wattenhofer, Dawn Song, and Arthur Gervais.
\newblock Sok: Decentralized finance (defi) attacks.
\newblock In {\em 2023 IEEE Symposium on Security and Privacy (SP)}, pages 2444--2461. IEEE, 2023.

\bibitem{zimperium2023Global}
Zimperium.
\newblock 2023 {G}lobal mobile threat report.
\newblock \url{https://get.zimperium.com/2023-global-mobile-threat-report/}, 2023.
\newblock [Accessed 08-03-2024].

\bibitem{securelistHelloName}
Konstantin Zykov.
\newblock {H}ello! my name is {D}track.
\newblock \url{https://securelist.com/my-name-is-dtrack/93338/}, 2019.
\newblock [Accessed 06-03-2024].

\end{thebibliography}

\end{document}